\documentclass[aps,prl,floatfix,twocolumn,superscriptaddress,twoside]{revtex4-2}
\usepackage[english]{babel}
\usepackage{booktabs}
\usepackage{graphicx}
\usepackage[colorlinks,citecolor=blue,linkcolor=blue,urlcolor=blue]{hyperref}
\usepackage{MnSymbol}
\usepackage{times}

\renewcommand{\Im}{\mathrm{Im}}
\renewcommand{\Re}{\mathrm{Re}}

\begin{document}

\title{Learning eigenstates of quantum many-body Hamiltonians within the symmetric subspaces using neural network quantum states}

\author{Shuai-Ting Bao}
\affiliation{School of Physics, Zhejiang University, Hangzhou 310058, China}

\author{Dian Wu}
\affiliation{Institute of Physics, École Polytechnique Fédérale de Lausanne (EPFL), CH-1015 Lausanne, Switzerland}

\author{Pan Zhang}
\affiliation{CAS Key Laboratory for Theoretical Physics, Institute of Theoretical Physics, Chinese Academy of Sciences, Beijing 100190, China}

\author{Ling Wang}
\email{lingwangqs@zju.edu.cn}
\affiliation{School of Physics, Zhejiang University, Hangzhou 310058, China}

\date{\today}

\begin{abstract}
  The exploration of neural network quantum states has become
  widespread in the studies of complicated quantum many-body
  systems. However, achieving high precision remains challenging due
  to the exponential growth of Hilbert space size and the intricate
  sign structures. Utilizing symmetries of the physical system, we
  propose a method to evaluate and sample the variational ansatz
  within a symmetric subspace. This approach isolates different
  symmetry sectors, reducing the relevant Hilbert space size by a
  factor approximately proportional to the size of the symmetry
  group. It is inspired by exact diagonalization techniques and the
  work of Choo et al.\ in Phys.\ Rev.\ Lett.\ {\bf 121}, 167204
  (2018). We validate our method using the frustrated
  spin-$\frac{1}{2}$ $J_1$-$J_2$ antiferromagnetic Heisenberg chain
  and compare its performance to the case without symmetrization. The
  results indicate that our symmetric subspace approach achieves a
  substantial improvement over the full Hilbert space on optimizing
  the ansatz, reducing the energy error by orders of magnitude. We
  also compare the results on degenerate eigenstates with different
  quantum numbers, highlighting the advantage of operating within a
  smaller Hilbert subspace.
\end{abstract}

\maketitle

\paragraph{Introduction.}
The advancement of neural quantum state
(NQS) has been impressive since its initial development, and numerous established models have been successfully applied to
solve challenges in quantum many-body systems.
Notable ones among these models include the convolutional
neural network (CNN)~\cite{Carleo_CNN19,Castelnovo20}, the graph neural network
(GNN)~\cite{Clark21,Li_google20,YuGNN_23,Ji_LCN_22}, the recurrent
neural network (RNN)~\cite{Carrasquilla20}, and the
transformer~\cite{Becca23_1,Becca23_2}, each demonstrating
remarkable achievements. The success of these models owes to their properties
that align closely with well-known physical concepts. For instance,
CNNs explicitly incorporate translational invariance, and the
graph attention mechanism in GNNs resembles the coarse-graining properties
of tensor network states~\cite{Verstraete04}. However, despite these
advancements, the complexity of quantum many-body wave functions
remains a formidable obstacle, attributed to the exponentially
large Hilbert space and the lack of information regarding their
sign structure.

Symmetry plays a crucial role in the design of variational ansatzes,
which rules out physically improper states and serves as a guidance in
variational optimization. Various methods have been proposed to
integrate the physical symmetries into neural networks. For example,
the CNN effectively captures the translational symmetry, but its
applicability is confined to ground state simulations in the trivial
symmetry sector, rendering it unsuitable for other symmetries such as
reflections, rotations, and sublattice translations. The group
convolutional network (GCN)~\cite{MacDonald21} explicitly preserves
the full symmetry group of the Hamiltonian. However,
  it faces scalability challenges because the size of its kernel
  (filter) increases in proportion to the size of the symmetry group,
  which itself grows with the system size. Discrete symmetries can
  also be enforced on the wavevector components by computing a
  phase-weighted average of the neural network outputs over the
  symmetry-transformed input
  states~\cite{Becca23_1,Becca23_2,Carleo_CNN19,Imada_PRX21,Feiguin_22},
  as $\psi(\sigma) = \sum_{g \in G} \chi_g
  \widetilde{\psi}(g^{-1}\sigma)$, where $g\in G$,
  $\widetilde{\psi}(\sigma)$ is a scalar output over input
  $\sigma$. The drawback of this approach is its computational
  expense. For any input $\sigma$, the calculation requires evaluating
  $\widetilde{\psi}(g\sigma)$ and its derivatives for all $g \in G$,
  making it inefficient for large groups.

There are several celebrated examples of incorporating symmetries
non-trivially into variational ansatzes. For instance, in a manner
analogous to the $\mathrm{SU}(2)$ symmetric matrix product state
(MPS), Vieira et al.~\cite{Verstraete20} use a fixed
  fusion routine (e.g., DMRG snake path) and assigns variational
  weights to specific chains of intermediate angular momenta
  $\{l_i\}$, where $i$ is the site index. While advantageous, this
  approach is demanding due to the need for Clebsch-Gordon
  coefficients and $ 6j$ symbols, making implementation cumbersome.

Choo et al.~\cite{Neupert18} explored translational symmetry by
ensuring that the amplitudes of real-space states related by
translations are identical while incorporating their relative phases
as dictated by the symmetry sectors. This method eliminates the
computational overhead of explicitly enumerating all group elements
and enables the investigation of low-energy excitations at different
momenta. Although represented in the full Hilbert space, the way it
imposes the symmetry is closely related to the construction of
symmetric subspace commonly used in exact diagonalization (ED) under
lattice symmetries and other conservation
laws~\cite{Laflorencie04,Noack05,Weisse08,Lauchli11,Sandvik10ed}.
Subsequently, Bukov et al.~\cite{Bukov_21} proposed a method that
confines the wavefunction to the symmetric subspace of the ground
state sector and perform the full sum in the subspace to guide the
optimization.

Inspired by both Ref.\cite{Neupert18} and
  Ref.\cite{Bukov_21}, this paper fully capitalizes on the advantages
  of symmetric subspace construction by applying both full summation
  and Metropolis sampling within the symmetric subspace, as opposed to
  traversing the entire Hilbert space~\cite{Neupert18}. This
methodology provides several key benefits: first, the inclusion of
symmetries increases the gap between the ground state and the lowest
excited state within the subspace, facilitating faster convergence
during variational optimization. Second, the non-local structure of
the symmetric basis vectors enhances the expressiveness of the
ansatz. Finally, the substantial reduction in Hilbert space size
enables the use of smaller neural networks and fewer Monte Carlo
samples, while achieving high accuracy.

\paragraph{Method.}
Similar to the symmetric basis construction in ED methods, we
construct a set of orthonormal basis vectors by superposing
symmetry-transformed Ising basis states, $|a\rangle = |s_1 \otimes s_2
\otimes \cdots \otimes s_N\rangle$, where $N$ is the system size, as
follows~\cite{Laflorencie04,Noack05,Weisse08,Lauchli11,Sandvik10ed}:
\begin{equation}
  |a_{\rm symm.}\rangle = \frac{1}{\sqrt{N_a}} \sum_{g \in G} \chi_g g^{-1} |a\rangle,
  \label{mbasis}
\end{equation}
where $g\in G$ represents an element of the symmetry group $G$, and
$\sqrt{N_a}$ is the normalization factor. The coefficient $\chi_g$ is
determined by the target quantum sector, ensuring that $g
|a_{\text{symm.}}\rangle = \chi_{g} |a_{\text{symm.}}\rangle$, so each
basis vector $|a_{\text{symm.}}\rangle$ is an eigenvector of the group
elements. A specific example of the translation-symmetric case could be 
\begin{equation}
  |a_{\rm symm.}\rangle = \frac{1}{\sqrt{N_a}} \sum_{r = 0}^{N - 1} e^{-i k r} T^r |a\rangle.
  \label{mbasis}
\end{equation}
The unique representative state $|a\rangle$ is selected as
the minimum one among the equivalence class $\{g |a\rangle\}$ for $g
\in G$ encoded in the binary format, with $1$ denoting
$|\!\uparrow\rangle$ and $0$ denoting
$|\!\downarrow\rangle$. $|a_{\text{symm.}}\rangle$
  corresponds one-to-one with its representative state $|a\rangle$,
  reducing the size of the Hilbert space by eliminating redundant
  basis $a' = ga$ that are linearly dependent. The symmetric
wavefunction $|\psi\rangle$ can be expanded as
\begin{equation}
  |\psi\rangle = \sum_{a \in \{\text{repr.\ state}\}}^{N_\text{repr}} \psi(a) |a_{\rm symm.}\rangle,
  \label{wavefunc1}
\end{equation}
where $N_\text{repr}$ is the size of the symmetric subspace. It is
significantly reduced compared to the full Hilbert space due to the
symmetry and the enforced orthogonality. The
  representative state $|a\rangle$, though part of the symmetric
  basis, is input into a neural network using its real-space
  configuration to retrieve the wavefunction coefficient $\psi(a)$.
In this framework, no extra computational effort is required beyond
enumerating or sampling representative configurations $a$, computing
the matrix elements $\langle a_{\rm symm.} | H | a'_{\rm symm.}
\rangle$, and evaluating the wave function coefficients $\psi(a)$. The
matrix elements in the symmetric subspace can be computed efficiently,
as usually adopted in ED
methods~\cite{Laflorencie04,Noack05,Weisse08,Lauchli11,Sandvik10ed}.

\begin{figure}
\centering
\includegraphics[width=0.8\linewidth]{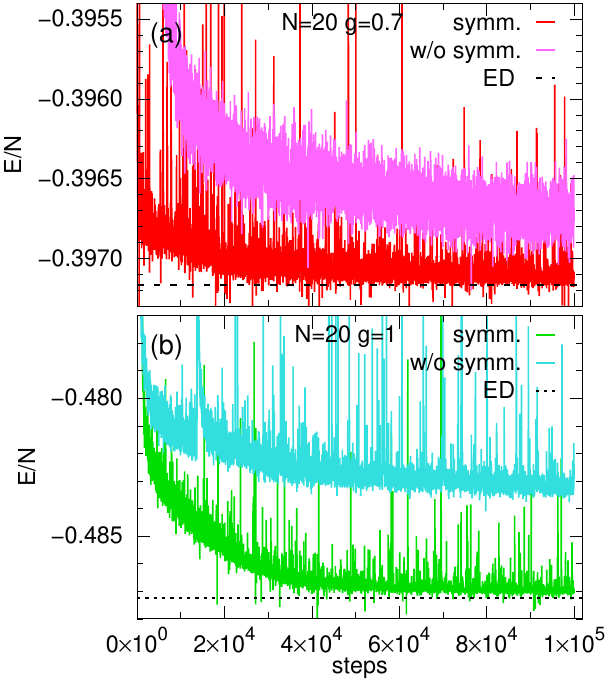}
\caption{Ground state energy vs.\ optimization steps for the
  $J_1$-$J_2$ chain of length $N = 20$ at (a) $g = 0.7$ and (b) $g = 1$,
  using Metropolis sampling. The NQS is a 4-layer complex FFNN with
  $\alpha_i = 2$. We compare the results with symmetrization to those
  with the same network and optimization strategy but without symmetrization.}
\label{enr-steps}
\end{figure}

\paragraph{Wave function ansatz.}
The architecture of NQS plays a crucial
role in the wave function ansatz. Apart from the abundant available models,
the methodologies for designing complex-valued wave functions are equally
significant. We explore two distinct approaches: employing a 
network with complex parameters, referred to as
ComplexNet throughout this paper; or employing two independent networks
with real parameters to model the amplitude and phase separately, referred to as
SeparateNet.

We use the most generic and effective neural network model, the
feed-forward neural network (FFNN), to implement the two approaches
and compare their accuracy and efficiency. The network
outputs the logarithm of the wave function components, written as
$\ln{\psi(a)} = \Psi(a)$. For ComplexNet, the output is a
complex number $\Psi(a)$, i.e.,
\begin{equation}
  \ln{\psi(a)} = \Re \Psi(a) + i \Im \Psi(a).
\end{equation}
For SeparateNet, the two real networks output $A(a)$ and $\Phi(a)$ respectively, and we have
\begin{equation}
  \ln{\psi(a)} = A(a) + i \Phi(a).
  \label{wavefunc}
\end{equation}
Formally, $e^{\Re \Psi(a)}$ (or $e^{A(a)}$)
represents the amplitude, and $\Im \Psi(a)$ (or $\Phi(a)$) represents
the phase of the component $|a_k\rangle$ for a ComplexNet
(or SeparateNet). For a FFNN with three layers as an example, we parameterize it as following:
\begin{equation}
\ln \psi(a) = \mathbf{W}_3 f\left( \mathbf{W}_2 f(\mathbf{W}_1 a + \mathbf{b}_1) + \mathbf{b}_2 \right),
\end{equation}
where $\mathbf{W}_i$ represents the weight matrix and $\mathbf{b}_i$
represents the bias, whose values are complex (or real) for ComplexNet
(or SeparateNet). The function $f$ is a nonlinear activation function,
for which we simply use the ReLU function in this work. The final weight matrix
$\mathbf{W}_3$ is a row vector, which outputs a scalar after the
matrix-vector multiplication. The number of hidden neurons at the
$i$-th layer is denoted as $N_{h_i} = \alpha_i N$. The complex ReLU
function is implemented by separately applying ReLU to its
real and imaginary parts. The energy gradients and the stochastic
reconfiguration (SR)~\cite{sorella98,sorella01,sorella07,becca17,Schmitt20} method to precondition the gradients are
presented in the supplementary material~\cite{supp}.

\paragraph{Metropolis sampling algorithm.}
To sample a given probability distribution
$p(a) = e^{2 \Re \Psi(a)}$ of configurations $a$, the Metropolis sampling must satisfy
the detailed balance condition~\cite{Greenberg19}:
\begin{equation}
p(a) g(a, a') = p(a') g(a', a),
\end{equation}
where $g(a, a')$ is the proposed transition probability from $a$ to
$a' \neq a$. It ensures that at equilibrium, the number of transitions
from $a$ to $a'$ equals the number of transitions from $a'$ to
$a$. We define $N_\text{tran}(a)$ as the number of
  distinct representative configurations $a'$ that are connected to
  $a$ via an operator $O$, where $O$ commutes with all symmetry
  operators $g$, i.e., $\langle a'_{\rm symm.} | O | a_{\rm symm.}
  \rangle \neq 0$. In this paper, we choose $O = \sum_{j=1}^2\sum_{i}
  (S^+_i S^-_{i + j} + S^-_i S^+_{i + j})$. Each $a'$ is proposed with
  equal probability, leading to $g(a, a') = 1 / N_\text{tran}(a)$.
The acceptance probability of the new representative configuration
$a'$ is then given by:
\begin{equation}
  r(a, a') = \min \left( 1, \frac{p(a') N_\text{tran}(a)}{p(a) N_\text{tran}(a')} \right).
\end{equation}

\begin{figure}
\centering
\includegraphics[width=0.7\linewidth]{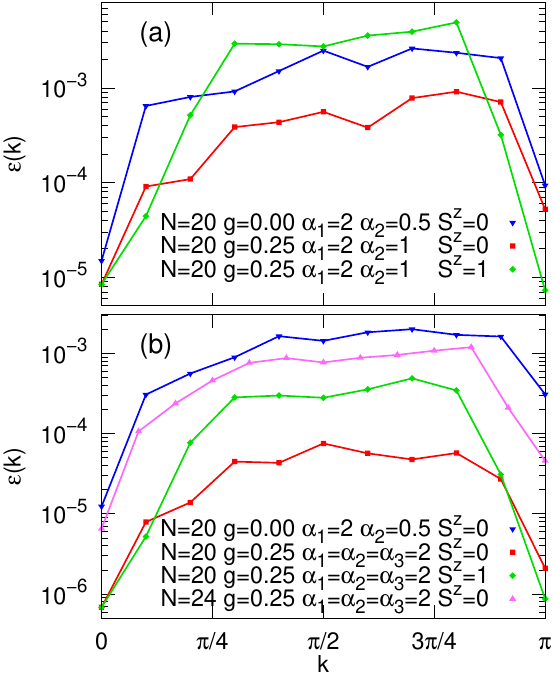}
\caption{Relative error in energy $\epsilon(k) = \frac{E(k) -
    E_\text{ED}(k)}{|E_\text{ED}(k)|}$ as a function of the momentum
  $k$, evaluated at $g = 0$ and $g=0.25$, within the symmetry sectors
  $S^z = 0$ and $S^z=1$, using full summation over the symmetric
  subspace.  (a) Results for $N = 20$ using a 3-layer SeparateNet. (b)
  Results for $N = 20$ and $N=24$ using 3-layer and 4-layer
  ComplexNets. The parameters $\alpha_i=N_{h_i}/N$
    indicates the ratio of hidden layer size to system size. Detailed
    explanations of the colored lines are provided in the main text.}
\label{err-momentum}
\end{figure}

\begin{figure}
\centering
\includegraphics[width=0.7\linewidth]{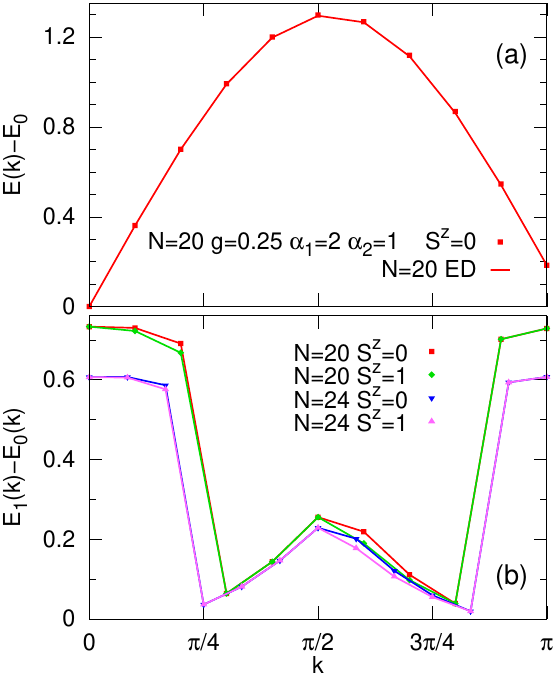}
\caption{(a) Spinon dispersion as a function of
    momentum $k$ at $g = 0.25$ for system size $N = 20$ and symmetry
    sector $S^z=0$, computed using a 3-layer SeparateNet with
    $\alpha_1=2$, $\alpha_2=1$, and full summation over the symmetric
    subspace, compared to ED. The corresponding energy relative error
    is shown in Fig.~\ref{err-momentum}~(a) in red squares. (b) The
    lowest non-degenerate energy gap, $\Delta(k) = E_1(k) - E_0(k)$,
    as a function of $k$, for system sizes $N = 20$ and $N=24$ at $g =
    0.25$, restricted to symmetry sectors $S^z = 0$ and $S^z=1$,
    computed by ED.}
\label{enr-momentum}
\end{figure}

\paragraph{Results.}
The Hamiltonian we select for demonstration is the frustrated
spin-$1/2$ $J_1$-$J_2$ antiferromagnetic Heisenberg model on the 1D chain with periodic boundary conditions, formulated as
\begin{equation}
  H = J_1 \sum_{\langle i, j \rangle} \mathbf{S}_i \cdot \mathbf{S}_j
  + J_2 \sum_{\llangle i, j \rrangle} \mathbf{S}_i \cdot \mathbf{S}_j,
  \label{hamilt}
\end{equation}
where the summations are performed over the nearest-neighbor pairs
$\langle i, j \rangle$ and the next-nearest-neighbor pairs
$\llangle i, j \rrangle$ respectively. Throughout this paper, we set $J_1 = 1$ and denote $g = J_2 / J_1$.
This simple model exhibits a diverse ground state phase
diagram~\cite{Nomura92,Eggert96,Neugebauer98}. Specifically, its
ground state is a Luttinger liquid (LL) for $g < g_{c 1} \approx 0.2411$, an
incommensurate spiral state for $g > g_{c 2} \approx 0.5294$, and a
valence bond solid (VBS) for $g_{c 1} < g < g_{c 2}$.

For $g=0$, the ground state satisfies Marshall’s sign
  rule: after applying a global spin rotation, $\prod_{i \in B}
  S^z_i$, on the $B$ sublattice, every component of the wavefunction
  becomes positive. For $g$ between 0 and $g_{c2}$, chains of length $N = 4n$
  strictly obey Marshall’s sign rule, whereas those with $N = 4n + 2$
  display only a small ($\sim 10^{-5}$ for $N\le 30$) deviation from the rule. Alternatively, the
Hamiltonian can be rotated as $H' = \prod_{i \in B} S^z_i H \prod_{i
  \in B} S^z_i$. Simulating $H^{\prime}$ ensures that the ground state
remains positive definite, simplifying the sign structure and
facilitating simulation and optimization with our numerical methods.

The Hamiltonian is invariant under the translational symmetry $T$, the
bond-mirror reflection symmetry $P$, and the spin inversion symmetry $Z =
\prod_i S^x_i$. We also incorporate the total spin conservation $S^z =
\sum_i S^z_i$. The ground state lies in the symmetry sector
with quantum numbers $k = 0$ (or $k = \pi$), $p = 1$ (or $p = -1$), $z = 1$
(or $z = -1$), and $S^z = 0$ for a chain of length $N = 4 n$ (or $N = 4 n +
2$), where $n$ is an integer, $p$ is the reflection quantum
number, and $z$ is the spin inversion quantum number. To compute its
low-energy dispersion when $k \neq 0$, we either set total spin $S^z = 1$, or set
$S^z = 0$ and $z = -1$ for $N = 4 n$ (or $z = 1$ for $N = 4 n + 2$).
We account for the reflection symmetry $P$ at
$k = \pi$ as well. The numbers of representative states $N_\text{repr}$ for various system sizes $N$
and quantum numbers $(k, p, z, S^z)$ are listed in the
supplementary material~\cite{supp}.

In Fig.~\ref{enr-steps}, we illustrate the ground state energy as a
function of optimization steps using Metropolis sampling for the $N =
20$ chain at couplings $g = 0.7$ and $g=1$, using a 4-layer ComplexNet
whose number of hidden neurons $N_{h_i} = \alpha_i N$, and we
configure $\alpha_i = 2$ for $i = 1, 2, 3$. The network has $4160$
complex parameters in total. The relative errors compared to the exact
ground state energy $E_0$ from ED, defined as $\epsilon = \frac{E -
  E_0}{|E_0|}$, are provided in
Table~\ref{err-size}. In this definition of
  $\epsilon$, $E$ is computed using full summation following the
  completion of the Metropolis sampling optimization procedure. As an
ablation study, we compare the results to those with the same network
and optimization strategy but without symmetrization, implemented
using the NetKet package~\cite{netketlib}. The symmetrization reduces
the energy relative errors by approximately two orders of
magnitude. Hyperparameters of the optimization are summarized in the
supplementary material~\cite{supp}.

\begin{table}
\centering
\begin{tabular}{c@{\hspace{1em}}cccc}
\toprule
& \multicolumn{2}{c}{$g = 0.7$} & \multicolumn{2}{c}{$g = 1$} \\
\cmidrule(lr){2-3} \cmidrule(lr){4-5}
& symm. & w/o symm. & symm. & w/o symm. \\
\midrule
$\epsilon$ & $6.3 \times 10^{-5}$ & $1.3 \times 10^{-3}$ & $1.7 \times 10^{-4}$ & $8.2 \times 10^{-3}$ \\
\bottomrule 
\end{tabular}
\caption{Relative errors of the ground state energy, defined as
  $\epsilon = \frac{E - E_0}{|E_0|}$, where $E_0$ represents the ED
  results, for the $J_1$-$J_2$ chain with $N = 20$ at $g = 0.7$ and
  $g=1$. The comparison highlights the results obtained with and
  without symmetrization. Here $E$ is computed using
    full summation following the completion of the Metropolis sampling
    optimization procedure.}
\label{err-size}
\end{table}

Next, we examine and compare the performances of SeparateNet and
ComplexNet using full summation over the symmetric subspace. We begin
by studying the low-energy dispersion as a function of the momentum
$k$ for a chain of length $N = 20$ at $g = 0$. We configure the ansatz
as a 3-layer SeparateNet (or ComplexNet) with $\alpha_1 = 2$,
$\alpha_2 = 0.5$, and $1260 \times 2$ real (or $1260$ complex)
parameters, restricted to the symmetry sectors $S^z = 0$ and $z =
-1$. The simulation accuracy is indicated by energy relative errors,
defined as $\epsilon(k) = \frac{E(k) -
  E_\text{ED}(k)}{|E_\text{ED}(k)|}$. The results with SeparateNet and
ComplexNet are presented as blue triangles in
Fig.~\ref{err-momentum}~(a) and (b) respectively. Both networks
achieve an accuracy on the order of $10^{-3}$, which is slightly
higher than the $3 \times 10^{-4}$ achieved in
Ref.~\cite{Neupert18}. This discrepancy is likely due to the parallel
tempering scheme used in Ref.~\cite{Neupert18}, which assists
optimization in reaching the global minimum. Regarding
  the comparison between SeparateNet and ComplexNet, the results show
  no discernible differences in performance. However, in terms of
  computational time, SeparateNet is twice as efficient when using the
  same network parameters (i.e., the same $\alpha_i$).

We study the low-energy dispersion at the coupling $g = 0.25$, which
is close to the critical point $g_{c_1} \approx 0.2411$ between the
Luttinger liquid (LL) and valence bond solid (VBS) phases. To explore
the effect of symmetric subspace size on variational energy, we
analyze degenerate states within different symmetry sectors. The
low-energy dispersion of this model, corresponding to the lowest
triplet excitations at momentum $k$, exhibits three-fold degeneracy
with quantum numbers $S^z = 0, \pm 1$.  For the $S^z = 0$ triplet, an
additional spin-inversion symmetry $Z$ applies, with quantum number $z
= -1$ for $N = 4 n$. Consequently, the subspace for the $S^z = 0$
triplet is approximately half that of the $S^z = \pm 1$ triplets.

To investigate the effect of subspace size, we employ a 3-layer
SeparateNet with parameters $\alpha_1 = 2$, $\alpha_2 = 1$, and $1680
\times 2$ real parameters on a chain of length $N = 20$ at $g =
0.25$. The relative errors $\epsilon(k)$ as a function of $k$ for the
$S^z = 0$ (in red squares) and $S^z = 1$
(green diamonds) sectors are presented in
Fig.~\ref{err-momentum}~(a). Away from from $k = 0$ and $\pi$, the
$S^z = 0$ sector achieves energy accuracy approximately an order of
magnitude better than the $S^z = 1$ sector, demonstrating the
advantage of a smaller variational subspace.  Near $k = 0$ and $\pi$,
both sectors achieve high accuracy, with the $S^z = 0$ sector showing
comparable performance to the $S^z = 1$ sector.

To understand the behavior near $k = 0$ and $\pi$, we compare the
spinon dispersion function $E(k) - E_0$ obtained from the neural
network and ED in Fig.~\ref{enr-momentum}~(a).  Additionally, the
lowest non-degenerate energy gap $\Delta(k) = E_1(k) - E_0(k)$,
calculated using ED, is presented in Fig.~\ref{enr-momentum}~(b)
for sizes $N=20$ and $24$ using the sectors $S^z=0$
  and $1$ respectively. The gap $\Delta(k)$ is particularly large
near $k = 0$ and $\pi$ for both sectors and sizes presented,
indicating better isolation of eigenstates within symmetric
subspaces. This isolation enhances the convergence accuracy and speed
of the variational optimization in these momentum regions.

We further enhance the model by employing a 4-layer ComplexNet with
$\alpha_i = 2$ and $4160$ complex parameters to compute the dispersion
for the $N = 20$ chain at $g = 0.25$. Results for the $S^z = 0$
(in red squares) and $S^z = 1$ (in
  green diamonds) sectors are presented in
Fig.~\ref{err-momentum}~(b). The increased number of variational
parameters significantly improves accuracy, reducing the relative
errors by over an order of magnitude compared to the 3-layer
SeparateNet results shown in Fig~\ref{err-momentum}~(a)
(in red squares and green diamonds respectively).

Lastly, we compute the relative error of the dispersion for the $N =
24$ chain at $g = 0.25$, constrained to the symmetry sector $S^z = 0$
with $z = -1$. Using a 4-layer ComplexNet with $\alpha_i = 2$ and
$5952$ complex parameters, the simulation achieves a maximum relative
error of approximately $1 \times 10^{-3}$, as
  illustrated by the magenta triangles in
  Fig.~\ref{err-momentum}~(b).

\paragraph{Discussion and conclusions.}
Building on previous advancements of exact diagonalization
(ED)~\cite{Laflorencie04,Noack05,Weisse08,Lauchli11,Sandvik10ed} and
symmetrization of neural quantum states (NQS)~\cite{Neupert18,Bukov_21}, we
have presented a method to evaluate and sample NQS within a significantly
reduced symmetric subspace, where all basis vectors are orthogonal,
non-local, and symmetric. By implementing a Metropolis sampling
algorithm that satisfies detailed balance, our method enables
large-scale simulations of symmetric NQS. Using
feedforward neural networks (FFNN) as the basic architecture, we have tested our approach on the
frustrated 1D $J_1$-$J_2$ antiferromagnetic Heisenberg
model. Our results demonstrate that, using the same network architecture and
optimization strategy to find the ground state
energy, a symmetric NQS achieves an accuracy approximately
two orders of magnitude better than a conventional NQS without
symmetry. Furthermore, simulations of low-energy
dispersion across different symmetry sectors reveal that a smaller
symmetric subspace significantly improves the variational energy, with
the relative error reduced by one order of magnitude when the subspace size is reduced by half.

This framework holds immense potential for the accurate simulation of
large-size many-body quantum Hamiltonians when combined with modern
neural network models such as graph neural networks
(GNN)~\cite{Clark21,Li_google20,YuGNN_23} and
transformers~\cite{Becca23_1,Becca23_2}. Its ability to simulate
excited states, in conjunction with spectroscopic
methods~\cite{Imada_PRX21,Wang18,Yang22,Wang22}, offers a promising
route for understanding subtle and intricate quantum phase
transitions. We look forward to an implementation that integrates the
symmetrization and the more efficient machine learning framework,
which will not only advance the precision of quantum state simulations
but also pave the way for exploring quantum phenomena in systems
larger than previously feasible.

\begin{acknowledgments}
We would like to thank G.\ Carleo for helpful comments. S.-T.B.\ and
L.W.\ are supported by the National Natural Science Foundation of
China (NSFC) (Grant No.~12374150).  P.Z.\ is supported by the NSFC
(Grant No.~12047503, 12325501, and 12247104) and Project
ZDRW-XX-2022-3-02 of the Chinese Academy of Sciences.
\end{acknowledgments}

\setcounter{equation}{0}
\setcounter{figure}{0}
\setcounter{table}{0}

\renewcommand{\theequation}{S\arabic{equation}}
\renewcommand{\thefigure}{S\arabic{figure}}
\renewcommand{\thetable}{S\arabic{table}}

\section*{END MATTER}
\vskip-2mm

\paragraph{Energy gradients.}
We denote the complex variational parameters as $\theta_c$ in
ComplexNet, and the real parameters in $A(a)$ and
$\Phi(a)$ as $\theta_r$ and $\theta_i$ respectively in SeparateNet. Given the
energy as the loss function, the gradient of energy with respect to
$\theta$ in general ($\theta_c$ for ComplexNet, or $\theta_r$ and
$\theta_i$ for SeparateNet) is expressed as:
\begin{equation}
\frac{\partial E}{\partial \theta} = \frac{\partial}{\partial \theta} \frac{\langle \psi | H | \psi \rangle}{\langle \psi | \psi \rangle}.
\end{equation}
After substituting $|\psi\rangle = \sum_a \psi(a) |a_k\rangle$, it becomes
\begin{align}
  \frac{\partial E}{\partial \theta} = &\phantom{{}+{}} \Big\langle \frac{\partial \ln \psi^*(a)}{\partial \theta} E_\text{loc}(a) 
  + \frac{\partial \ln \psi(a)}{\partial \theta} E^*_\text{loc}(a) \Big\rangle_a \nonumber \\
  &- 2 \langle H \rangle_a \Big\langle \frac{\partial \Re \ln \psi(a)}{\partial \theta} \Big\rangle_a,
\end{align}
given that $\langle O \rangle_a = \frac{\sum_a |\psi(a)|^2 O(a)}{\sum_a |\psi(a)|^2}$,
$E_\text{loc}(a) = \sum_{a'} \frac{\psi(a')}{\psi(a)} H_{a, a'}$,
and $H_{a, a'} = \langle a_k | H | a'_k \rangle$.
For ComplexNet, we have
\begin{align}
  \nonumber
  \frac{\partial E}{\partial \theta_c} = &\phantom{{}+{}} 2 \Big\langle \frac{\partial \Re \Psi(a)}{\partial \theta_c} \Re E_\text{loc}(a) \Big\rangle_a
  - 2 \langle H \rangle_a \Big\langle \frac{\partial \Re \Psi(a)}{\partial \theta_c} \Big\rangle_a \nonumber \\
  &+ 2 \Big\langle \frac{\partial \Im \Psi(a)}{\partial \theta_c} \Im E_\text{loc}(a) \Big\rangle_a.
  \label{complexderi}
\end{align}
Alternatively, for SeparateNet,
\begin{align}
  \frac{\partial E}{\partial \theta_r} &= 2 \Big\langle \frac{\partial A(a)}{\partial \theta_r} \Re E_\text{loc}(a) \Big\rangle_a
  - 2 \langle H \rangle_a \Big\langle \frac{\partial A(a)}{\partial \theta_r} \Big\rangle_a, \nonumber \\
  \frac{\partial E}{\partial \theta_i} &= 2 \Big\langle \frac{\partial \Phi(a)}{\partial \theta_i} \Im E_\text{loc}(a) \Big\rangle_a.
  \label{realderi}
\end{align}
Eq.~\ref{complexderi} and Eq.~\ref{realderi} bear resemblance to each
other, thus exhibiting similar computational complexity. We have compared and discussed
their performances in the main text.

Given the gradients obtained above, we further use stochastic
reconfiguration (SR)~\cite{sorella98,sorella01,sorella07,becca17,Schmitt20} to
regularize the optimization step for each variable. Following
Refs.~\cite{Neupert18,Schmitt20}, we consider updating the wavefunction
by an infinitesimal step $\delta \theta$:
\begin{equation}
  \psi(\theta + \delta \theta) = \psi(\theta) + \sum_\theta \delta \theta \frac{\partial \psi}{\partial \theta} 
  = \psi(\theta) \left( 1 + \sum_\theta \delta \theta \frac{\partial \ln \psi}{\partial \theta} \right).
\end{equation}
The best solution $\delta \theta$ given by SR should minimize the distance between
$\psi(\theta + \delta \theta)$ and the imaginary-time evolved
wavefunction
\begin{equation}
  \tilde{\psi}(\theta) = e^{-\tau H} \psi(\theta) = (1 - \tau H) \psi(\theta).
\end{equation}
Their distance is defined as
\begin{equation}
  d = \arccos \left( \sqrt{\frac{\langle \psi(\theta + \delta \theta) | \tilde{\psi}(\theta) \rangle
  \langle \tilde{\psi}(\theta) | \psi(\theta + \delta \theta) \rangle}
  {\langle \tilde{\psi}(\theta) | \tilde{\psi}(\theta) \rangle
  \langle \psi(\theta + \delta \theta) | \psi(\theta + \delta \theta) \rangle}} \right).
\end{equation}
Let's define $d = \arccos \sqrt{1 - x}$.
Using Taylor expansion $\arccos\sqrt{1 - x} \approx \sqrt{x}$ for small
$x$, and keeping up to the second order in $\tau$ and $\delta \theta$, we
obtain a simplified $x$ as follows:
\begin{equation}
  x \approx \sum_{ij} \delta \theta^*_i S_{i, j} \delta \theta_j 
  + \tau (\sum_i \delta \theta^*_i F_i + \sum_i F^*_i \delta \theta_i)
  + \tau^2 (\langle H^2 \rangle - \langle H \rangle^2),
  \label{distancesquare}
\end{equation}
where $S_{i, j} = \langle \frac{\partial \ln \psi^*}{\partial \theta_i} \frac{\partial \ln \psi}{\partial \theta_j} \rangle
  - \langle \frac{\partial \ln \psi^*}{\partial \theta_i} \rangle \langle \frac{\partial \ln \psi}{\partial \theta_j} \rangle$,
and $F_i = \langle \frac{\partial \ln \psi^*}{\partial \theta_i} E_\text{loc} \rangle 
  - \langle \frac{\partial \ln \psi^*}{\partial \theta_i} \rangle \langle H \rangle$.
Minimizing Eq.~\ref{distancesquare} with respect to
$\delta \theta^*_i$ results in the linear equations
\begin{equation}
  S_{i, j} \delta \theta_j = -\tau F_i,
  \label{s9}
\end{equation}
which can be solved by directly inverting the geometric matrix $S$, or
by an iterative solver. For ComplexNet, the expression for $F_i$ is
\begin{align}
  \nonumber
  F_i = &\phantom{{}+{}} \Big\langle \frac{\partial \Re \Psi(a)}{\partial \theta_c} \Re E_\text{loc}(a) \Big\rangle_a
  - \Big\langle \frac{\partial \Re \Psi(a)}{\partial \theta_c} \Big\rangle_a \Big\langle \Re E_\text{loc}(a) \Big\rangle_a \\
  \nonumber 
  &+ \Big\langle \frac{\partial \Im \Psi(a)}{\partial \theta_c} \Im E_\text{loc}(a) \Big\rangle_a
  - \Big\langle \frac{\partial \Im \Psi(a)}{\partial \theta_c} \Big\rangle_a \Big\langle \Im E_\text{loc}(a) \Big\rangle_a \\
  \nonumber
  &-i \bigg( \Big\langle \frac{\partial \Im \Psi(a)}{\partial \theta_c} \Re E_\text{loc}(a) \Big\rangle_a
  - \Big\langle \frac{\partial \Im \Psi(a)}{\partial \theta_c} \Big\rangle_a \Big\langle \Re E_\text{loc}(a) \Big\rangle_a \bigg) \\
  &+i \bigg( \Big\langle \frac{\partial \Re \Psi(a)}{\partial \theta_c} \Im E_\text{loc}(a) \Big\rangle_a
  - \Big\langle \frac{\partial \Re \Psi(a)}{\partial \theta_c} \Big\rangle_a \Big\langle \Im E_\text{loc}(a) \Big\rangle_a \bigg),
\end{align}
while for SeparateNet, $F_i$ is simply $[\frac{\partial E}{\partial
  \theta_r}, \frac{\partial E}{\partial \theta_i}]$, where the brackets
denote concatenation.

\paragraph{Sizes of symmetric subspaces.}
We summarize the sizes of the symmetric subspaces for the ground states
and their low energy dispersions for chain lengths $N = 16$, $20$, and
$24$ in Table~\ref{nrepr-size}. The ground states have the highest
symmetry, with the quantum number set $(k, p, z, S^z) = (0, 1, 1, 0)$ for
system sizes $N = 4 n$. The lowest triplet gap locates at $(k, p, z, S^z) = (\pi, -1, -1, 0)$
and $(k, p, z, S^z) = (0, 1, \text{N/A}, \pm 1)$.
Note that the unusual momentum quantum number $k = 0$ and
reflection quantum number $p = 1$ for the lowest triplet with $S^z = \pm 1$
are caused by the global spin rotation $\prod_{i \in B} S^z_i$ on
the sublattice $B$. Under the original Hamiltonian, the quantum number set
would be $(k, p, z, S^z) = (\pi, -1, \text{N/A}, \pm 1)$. At other
momenta $k \neq 0$ or $\pi$, the reflection symmetry is not
applicable.

\begin{table*}
\centering
\begin{tabular}{|c|c|c|c|c|c|}
\hline
~$N$~ & $(0, 1, 1, 0)$ & $(\pi, -1, -1, 0)$ & $(0, 1, \text{N/A}, \pm 1)$ & $(\pi/2, \text{N/A}, -1, 0)$ & $(\pi/2, \text{N/A}, \text{N/A}, \pm 1)$ \\
\hline
~16~ & 257 & 230 & 375 & 396 & 715 \\
\hline
~20~ & 2518 & 2429 & 4262 & 4639 & 8398 \\
\hline
~24~ & 28968 & 28648 & 52234 & 56275 & 104006 \\
\hline 
\end{tabular}
\caption{Sizes of the symmetric subspaces for the ground states (Column
  2), the lowest triplet states (Columns 3 and 4), and at $k = \pi/2$
  (Columns 5 and 6), denoted by their quantum number sets $(k, p, z,
  S^z)$, for system sizes $N = 16$, $20$, and $24$ respectively.}
\label{nrepr-size}
\end{table*}

\begin{figure}
\centering
\includegraphics[width=0.7\linewidth]{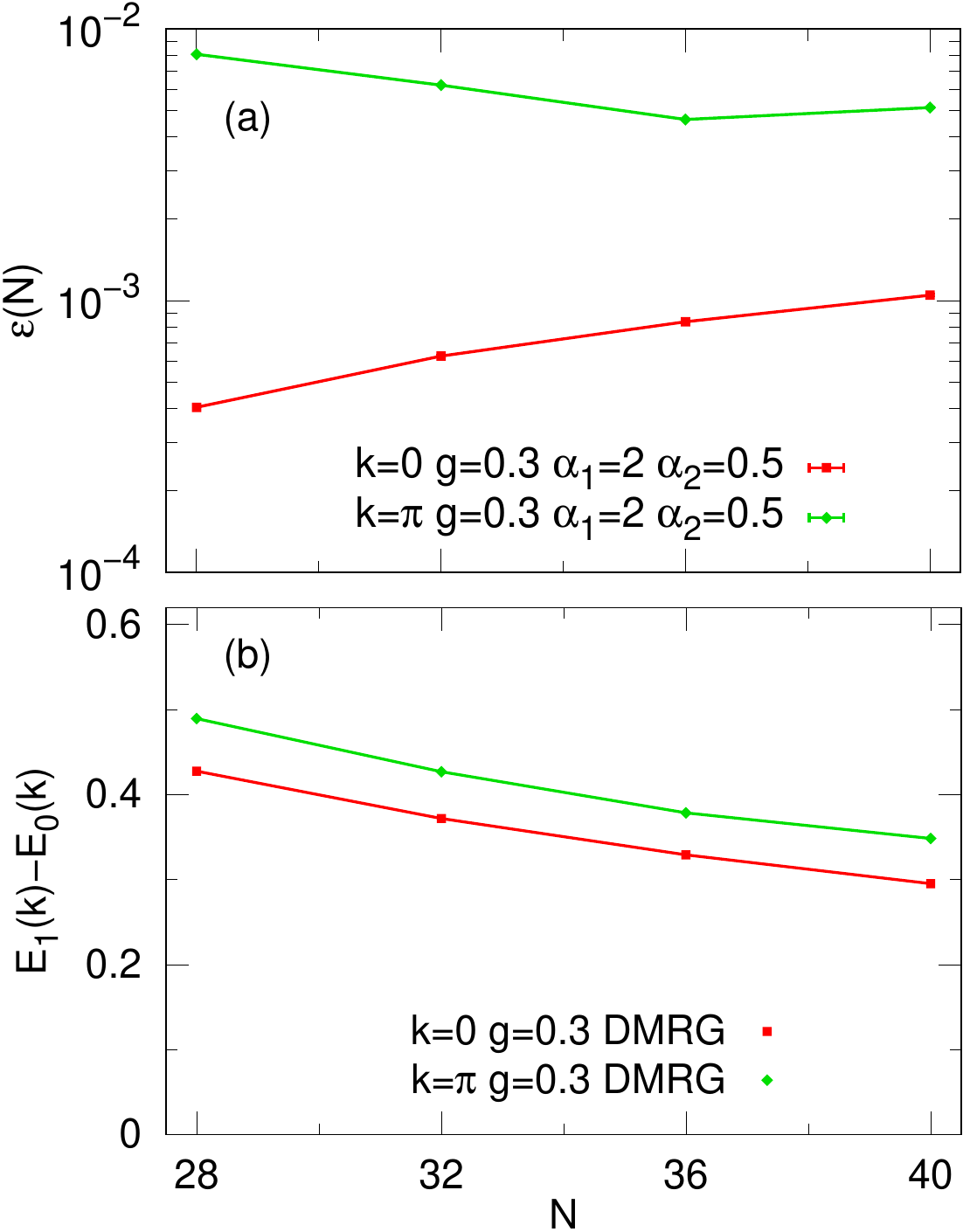}
\caption{(a) Relative energy errors $\epsilon(k) = \frac{E(k) - E_{\text{DMRG}}(k)}{|E_{\text{DMRG}}(k)|}$ (Eq.\ref{errork}) as a function of system size $N=28,32,36,40$ at $g=0.3$ for $k=0$ and $k=\pi$, obtained using a three-layer FFNN with $\alpha_1=2$ and $\alpha_2=0.5$. (b) Energy gap $E_1(k) - E_0(k)$ (Eq.\ref{gapk}) within each quantum sector, computed via DMRG, as a function of $N$ for $k=0$ and $k=\pi$.}
\label{benchmarklargeL}
\end{figure}

\paragraph{Optimization strategy.}
In the Monte Carlo (MC) simulation in the main text, we examine two typical
couplings at $g = 0.7$ and $1$. For the system size $N = 20$, we set the
number of samples per optimization step $n_\text{sample} = 4800$, the number of
Markov chain steps to collect the samples $n_\text{collect} = 2 N$, and the
number of steps to discard before collecting the samples $n_\text{discard} = 10 N$. We choose the variational wavefunctions to be 4-layer FFNN with $\alpha_i=2$ for each $i$ because of its complex sign structure.

  To address the efficiency of the Monte Carlo sampling procedure, we tested it on a relatively large system with $28 \le N \le 40$ using a less frustrated setup at $g=0.3$. Using exact diagonalization (ED), we verified that within the parameter range $0<g<g_{c2}$, chains of length $N=4n$ strictly obey Marshall’s sign rule, whereas those with $N=4n+2$ display only a small deviation from the rule.

For our variational trial wavefunction, we employed a
  three-layer feedforward neural network (FFNN) with $\alpha_1=2$ and
  $\alpha_2=0.5$, targeting the lowest eigenstates at $k=0$ and
  $k=\pi$. The sampling parameters were set as follows: the number of
  samples was $N_{\text{sample}}=12800$, the number of Markov chain
  steps used for sample collection was $n_\text{collect} = 1$, and the
  number of steps discarded before collection was $n_\text{discard} =
  1$. In addition to the benchmark examples discussed in the main
  text, we employed a parallel tempering scheme with a uniform inverse
  temperature spacing and $N_{\text{temp}}=10$, similarly defined as
  in Ref.~\cite{Neupert18}. A key difference between the large-system
  benchmark presented here and the smaller-system benchmark in the
  main text is that, with $n_\text{collect} = 1$, essentially no
  Markov chain samples are discarded. This is made possible by the
  equilibration effect provided by the parallel tempering scheme,
  ensuring that the sampling process remains efficient with no
  auto-correlation between samples.

The results for the relative energy errors, defined as
\begin{equation}
  \label{errork}
\epsilon(k) = \frac{E(k) - E_{\text{DMRG}}(k)}{|E_{\text{DMRG}}(k)|}
\end{equation}
as a function of system size $N$, are shown in Fig.~\ref{benchmarklargeL}(a). Here, $E_{\text{DMRG}}(k)$ represents the ground-state energy in momentum sector $k$, computed using an SU(2) DMRG algorithm with carefully converged bond dimensions $D=50,100,200$ for $N=28,32,36,40$. We observe that the relative energy error is on the order of $10^{-4}$ for states in the $k=0$ sector and on the order of $10^{-3}$ for states in the $k=\pi$ sector, highlighting the increasing complexity of wavefunctions for excited states.

To better understand how these errors are related to the energy gaps within each sector, we also examined the energy gaps, defined as
\begin{equation}
  \label{gapk}
\Delta_k = E_1(k) - E_0(k),
\end{equation}
where $E_1(k)$ (the first excited state) and $E_0(k)$ (the ground state) were computed using the DMRG algorithm with the ground-state projection method. As shown in Fig.~\ref{benchmarklargeL}(b), both gaps remain finite and relatively large. This implies that the primary source of the energy accuracy difference between the $k=0$ and $k=\pi$ states arises from variations in energy density at these momentum values.

For a direct numerical comparison at $N=30$ and $g=0.3$, Ref.~\cite{becca22} reports a relative energy error of $10^{-5}$ (Fig. 12 of Ref.~\cite{becca22}), wherea our best result (Fig.~\ref{benchmarklargeL}(a)) is approximately $5\times 10^{-4}$. The larger error in our case indicates that our optimization did not reach its global minimum. We believe this is primarily due to (i) the increased difficulty of optimizing a higher-dimensional parameter space, and (ii) our choice not to employ Stochastic Reconfiguration (SR) because it increases computational time by a factor of 10–20.

We utilize the Adam optimizer~\cite{adampaper} with the first momentum $\beta_1 = 0.9$ and
the second momentum $\beta_2 = 0.999$ throughout this work.
When optimizing from a random initialization, we set the learning rate $\alpha =
10^{-3}$. When fine-tuning from a converged NQS, $\alpha = 10^{-5}$ is
used instead. For MC sampling, we do not employ the SR method due to its
time-consuming nature. For optimization with full summation, the SR
method is employed for comparison.
To prevent singular or very small eigenvalues of the geometric
matrix $S$, we add an diagonal shift of size $10^{-3}$ to it. The
linear equations are solved using the SciPy~\cite{scipy} function
\texttt{gmres}, with the maximum number of iterations set to $40$. The initial
vector of the \texttt{gmres} function is always set to be $F_i$ in
Eq.~\ref{s9}.

\end{document}